\documentstyle[12pt,aaspp4]{article}

\newcommand{\etal}{{\it et al.}}

\begin{document}

\title{The S2 VLBI Correlator:\\
       A Correlator for Space VLBI and Geodetic Signal Processing}
\author{B. R. Carlson, P. E. Dewdney, T. A. Burgess, R. V. Casorso}
\affil{National Research Council of Canada,
       Herzberg Institute of Astrophysics, \\
       Dominion Radio Astrophysical Observatory, \\
       Penticton, British Columbia, Canada \\
       Electronic mail: Brent.Carlson@hia.nrc.ca,
       Peter.Dewdney@hia.nrc.ca}
\authoraddr{DRAO/HIA/NRC, \\
            P.O. Box 248, \\
            Penticton, B.C. \\
            Canada V2A 6K3 \\
            Tel: (250) 493-2277}

\author{W. T. Petrachenko}
\affil{Natural Resources Canada,
       Dominion Radio Astrophysical Observatory, \\
       Penticton, British Columbia, Canada \\
       Electronic mail: Bill.Petrachenko@hia.nrc.ca}

\author{and W. H. Cannon}
\affil{Space Geodynamics Laboratory,
       Center for Research in Earth and Space Technology, \\
       4850 Keele Street, North York, Ontario, Canada \\
       Electronic mail: wayne@sgl.crestech.ca}


\begin{abstract}

A unique lag-based VLBI correlator system has been developed for the
purpose of supporting both S2-based Space VLBI observations in the
Japanese-led VSOP mission and the Canadian Geodetic VLBI program. The
system architecture has been designed so that replication of a small
number of modules can be used to construct systems with a wide range
of sizes.  Optimized for a large correlator, the design is
`station-based' in the sense that as many hardware and software
functions as possible are performed before data is replicated and
transmitted for baseline (station pair) processing. As well as delay
compensation and generation of phase rotation coefficients,
station-based functions include autocorrelation, tone extraction,
pulsar gating, signal-statistics accumulation, and digital
filtering. Doppler shift correction (fringe stopping) is performed on
a baseline basis {\it at each correlator lag} so that there are no
smearing effects (lag-dependent loss of coherence) or frequency shifts
that must otherwise be corrected after correlation.  This is a key
element that simplifies the baseline processing architecture when high
accelerations associated with an orbiting antenna must be
considered. Flexible, efficient distribution of data from
station-based hardware to baseline-based hardware is accomplished by
serializing the wide data paths to 1 Gbit/s signals and using
high-speed switches to route the signals to their final destinations
where they are de-serialized before cross-correlation. This greatly
reduces the size, wiring complexity, and cost of the system. The
interval between updates of the delay models, integration times, and
other important events is typically 10 ms, but can be as short as 1
ms. Within this period, delay and fringe model generation is performed
using linear hardware synthesizers. The correlator also contains a
number of unique signal processing functions that extend its
capability beyond a basic VLBI correlator: flexible Local Oscillator
frequency switching for bandwidth synthesis; rapid (1 ms) correlator
dump intervals (allowing, for example, the study of some single-pulse
pulsar characteristics on VLBI baselines); simple but powerful
multi-rate digital signal-processing techniques to allow correlation
of signals at different but related sample rates; and a digital `zoom'
filter for producing very high resolution cross-power spectra. The
correlator software, written almost entirely in `C', is highly
integrated into the system, supports all of the functions mentioned
above, and is re-configurable to support expansion of the correlator.
The software schedules the use of hardware resources to enable
correlation of multiple observations concurrently, and automatically
schedules the correlation of observations that require more than the
available number of physical playback terminals. There is also
substantial pre-correlation consistency checking. The delay model is
based on CALC for ground-based antennas and NAIF for space-based
antennas. Output data is stored in the UVFITS format. The paper
describes the design rationale, architecture and function of the
correlator, and also provides specifications for the implemented
system.

\end{abstract}



\section{Introduction}

Very Long Baseline Interferometry (VLBI) is a well established
technique used to achieve milliarcsecond resolution with cm wave radio
telescopes operating in an interferometer array configuration on a
continental or global basis.  Large separation between interferometer
elements is desirable, since as the separation increases, the
resolution of the resulting image increases.  In VLBI, due to the
distance between interferometer elements and the large bandwidths
involved, it is not currently practical to combine the signals from
these elements in real time.  Thus wide-band recording technology has
been developed so that the signals may be recorded at observing time
and combined or correlated at a central site some time later.  Most
notably, the MkIII \markcite{Whitney1982} (Whitney 1982) and VLBA
\markcite{Hinteregger1991} (Hinteregger \etal\ 1991)
\markcite{Napier1994} (Napier \etal\ 1994) observing, recording, and
correlation systems have been in widespread use for a number of years.
The MkIII is now being replaced with a much higher performance MkIV
system and the VLBA has seen performance increases since its initial
design. Recently, a new low-cost S2 recording system
\markcite{Wietfeldt1996}(Wietfeldt \etal\ 1996) and correlation system
have been developed in Canada for the purpose of supporting Space VLBI
and geodetic VLBI \markcite{Cannon1997} (Cannon \etal\ 1997). The S2
correlator, described herein, can also be upgraded with recorders up
to four times the bandwidth of the S2 recorders.  The S2 recording
system is a descendent of the S1 system \markcite{Yen1988}(Yen \etal\
1988). The main reason for developing this system has been to provide
a low-cost recording and playback system that meets VSOP mission
requirements, thus enabling a more diverse set of ground radio
telescopes to participate in space and other VLBI programs than would
be available if the MkIV or VLBA systems were used exclusively.

Space VLBI overcomes the fundamental limit imposed on ground-based
VLBI -- the size of the Earth -- by placing a parabolic antenna into
Earth orbit. Thus, separations between antennas ({\em i.e.} baselines)
several times the diameter of the Earth can be achieved. The orbit is
chosen to be highly eccentric so that many different spacings between
interferometer elements can be obtained during the course of an
observation, thus producing a synthesized aperture
\markcite{Thompson1986} (Thompson \etal\ 1986) that is as completely
filled as possible.  Although the concept is straightforward, there
are complicating factors which make the operation of a Space VLBI
system more complex than a ground-based system.  Aspects of this which
affect correlator design are: correction for variations in the clock
for the Space Radio Telescope (SRT)\footnote{VLBI antennas on the
ground synchronize to local hydrogen-maser time standards.  Since
construction of reliable and cost-effective space-qualified hydrogen
masers for Space VLBI are still under development in Switzerland and
elsewhere, a clock signal is transmitted to the SRT over a telemetry
link to the spacecraft (see also \markcite{Springett1992} (Springett
1992)).}, and the ability to track delay and Doppler shift from an
antenna with a velocity as high as 10 km/sec and an acceleration close
to 1 g.  The Japanese-led VSOP (VLBI Space Observatory Programme)
\markcite{Hirosawa1996} (Hirosawa and Hirabayashi 1996) mission is
currently in the fully operational state and is successfully making a
few dozen observations per month with the HALCA satellite that was
launched on February 12, 1997 \markcite{Hirabayashi1998} (Hirabayashi
\etal\ 1998).

The correlator described in this paper has been specifically designed
to handle the long baselines, and the high velocities and
accelerations of Space VLBI.  It is now one of three operational Space
VLBI correlators and processes data from up to thirteen S2-equipped
ground radio telescopes and five S2-equipped tracking stations on a
regular basis --- mostly from the VSOP mission. In addition, it was
designed to process frequency-switched observations used in the
geodetic VLBI application to accurately measure delay, using the
`bandwidth synthesis' technique \markcite{Petrachenko1993}
(Petrachenko \etal\ 1993).

VLBI correlators are devices whose complexity rivals that of entire
telescopes. They are typically operated as facilities for a wide
variety of astronomical and geodetic observers. The purpose of this
paper is to describe in some detail the design rationale,
architecture, and function of this correlator. The paper is directed
both at the prospective builders of future VLBI correlators and at the
sophisticated user of this correlator.


\section{Hardware System Architecture} The underlying principle
guiding the design of the correlator system hardware was: {\it
Minimize the amount of hardware (and resulting software) design work
and maximize the functionality, flexibility, and expandability of the
system}.  A lag-based (`XF'\footnote{An `XF' correlator performs
time-domain cross-correlation and integration followed by Fourier
transform to get the final result.  An `FX' correlator performs a
real-time Fourier transform followed by multiplication and
accumulation to get the final result.  In principle, there are fewer
operations in the FX correlator, but the operations that must be
performed after Fourier transform must use longer word lengths than
those performed in a 1 or 2-bit XF correlator.  FX correlators have
been successfully built but are, in the opinion of the designers of
this correlator, more complex and difficult to implement than XF
correlators.}) correlator architecture was chosen and wherever
possible, industry-standard circuit board form factors, connectors,
bus interfaces, and protocols were used.  This resulted in the design
of a small number of modules, each with well defined interfaces and
functionality that can be used as building blocks to construct
virtually any size system.  Fig.~\ref{fig1} is a simplified block
diagram which shows a multi-station correlator including all of the
important building blocks --- the S2-Playback Terminal (S2-PT),
Station Module, Serial Distribution Module (SDM), and Baseline Module.

As with most digital systems, many performance characteristics are
dependent upon the timing of computer-generated events. In this paper
the minimum interval between events is called the `correlator
time-slice', and governs the frequency with which delay and phase
models can be updated, data read out, etc.  In the system described
here, the correlator time-slice can be either 10 ms or 1 ms. More
detail on this subject is presented in sections~\ref{PGFD}
and~\ref{builtsystem}.


\begin{figure}
\vspace{14cm}

\caption[]{
Simplified correlator block diagram showing
all critical modules.  Data originates from the S2-PT, undergoes
station-based processing in the Station Module, and is distributed by
Serial Distribution Modules (SDMs) to Baseline Modules where
correlation occurs.  Embedded CPUs controlling hardware and reading
out data are commanded by one or more Host Computers.
}
\label{fig1}
\end{figure}



\subsection{S2 --- VLBI Playback Terminal (S2-PT)} The S2-PT is a
highly automated, self-contained tape machine that plays back sampled
baseband data\footnote{In radio astronomy antennas, data is sampled
or quantized to 1 or 2 bits per baseband channel.} at 4, 8, 16, or 32
Msample/sec previously recorded
on an S2-Record Terminal (S2-RT) at a receiving antenna site. The
control and data interface to the S2-PT is a functionally complete,
well-defined interface that minimizes the amount of tape-specific
processing needed in the correlator. The S2-PT uses VHS recording
technology, modified to record digital data, and additional control
electronics to provide the correlator with up to 16 bit-streams of
digital data in its original sampled form along with precise timing
information.  These data include data-valid flags that indicate
whether samples from the tapes are good or bad.  Timing information is
provided to the correlator via a 1 Hz time-tick in conjunction with a
dedicated communications link called the Recorder Control Link
(RCL). The correlator automatically controls tape motion by sending
high-level commands to the S2-PT on the RCL.  The S2-PT responds to
commands and informs the correlator of its current state.


\subsection{Station Module} The Station Module (Fig.~\ref{fig2}) maps
the bit-streams from the S2-PT (hereafter referred to as simply `the S2')
into up to eight channels of 2-bit
quantized data and data-valid bits and performs all required
station-based processing for a single station.  This includes station
delay compensation to $\pm0.5$ samples of delay, per channel 4-bit
quantized fringe phase generation to be used in downstream
cross-correlation, autocorrelation, phase calibration tone
extraction\footnote{Many VLBI antennas inject a tone comb across the
received band --- extraction of the amplitude and phase of these tones
permits fault diagnosis and calibration of the phase and amplitude in
the path from the antenna feeds to the sampler
(Section~\ref{ToneExtract}).}, and quantizer state histogram accumulation. The
Station Module is implemented with a number of in-system programmable
Field Programmable Gate Arrays (FPGAs) and can thus be reprogrammed
to either correct functional problems or implement new
functionality within the constraints of the existing hardware
connectivity.  The Station Module is implemented on a 9U x 400 mm
standard VME circuit board. (The VME standard is described in
ANSI/IEEE STD1014-1987, IEC 821 and IEC 297, published by the IEEE
Standards Board and the American National Standards Institute.)


\begin{figure}
\vspace{14cm}

\caption[]{
Simplified Station Module block diagram
showing all important functional blocks.  Data enters the module from
the S2 via the P3 connector.  The main flow of data is shown by the
dark lines through the middle of the diagram.  The P2 connector is the
port for the standard VSB bus which is used to control and read out
data from the module using an external CPU board.  The P1
connector is used only for routing the system-wide clock onto the
module.  The High Speed Serial Multiplexer block converts parallel
data to one or two high speed bit streams for transmission to downstream
correlator modules.
}
\label{fig2}
\end{figure}


\subsubsection{\label {DG} Delay Generator} The function of the delay
model generator is to produce a value of real-time station delay at
each sample time.  There is one delay generator for each Station
Module. It contains a 32-bit delay accumulator and a 32-bit delay rate
register that are initialized with new coefficients every correlator
time-slice and otherwise are free-running to yield a delay point every
sample time.  Station-based delay tracking is accomplished in hardware
by looking for integer-sample delay changes in the delay accumulator
and adjusting the main FIFO read clock accordingly. This aligns the
data coming out of the Station Module to within $\pm0.5$ samples of
delay.  In addition, the upper 10 bits of the delay accumulator are
serialized and sent along with the data (restricting the delay update
period to once every 10 sample clock times) to the downstream
correlator modules where baseline-based fine delay tracking and phase
modification are performed.  The integer part of the 10-bit word is
fundamental to the operation of the Phase Modifier (see
section~\ref{VDPM}) and the remaining bits are used to minimize the
error incurred in the difference occuring in
equations~(\ref{equation1}) and (\ref{equation2}).  The 10-bit word
contains 0, 2, 3, or 4 bits of integer-sample delay information and
10, 8, 7, or 6 bits, respectively, of fractional-sample delay
information, corresponding to values of the sampling factor, $SF$, of
$\infty$, 4, 8, and 16 respectively. ($SF$ and the derivation of the
baseline based delay tracking are described in section~\ref{VDPM}.) In
this implementation, there are no practical limitations to the delay
rate that can be tracked.

\subsubsection{\label {FG} Fringe Generators} The function of the
fringe generator is to calculate values used to de-rotate
interferometer phase, which are then transmitted to the Baseline
Modules to be applied to the data, as described in section~\ref{CCA}.
One independently programmable fringe-phase generator is available for
each of the eight data channels.  Each generator contains a 32-bit
phase accumulator and a 32-bit phase rate register that, like the delay
generator, is initialized with new coefficients every correlator time-slice
and is otherwise free-running to yield a phase every sample time.  At
every sample clock time, the upper 4 bits of each phase accumulator are
effectively compressed to one serial bit-stream each, and sent along with
the associated channel's data, to the Baseline Modules.  (This is
implemented by sending the upper 4 bits of the phase and phase-rate
coefficients to the Baseline Modules every correlator timeslice, and thereafter
sending the carry bit from the lower 28-bits of the phase register to the
Baseline Modules every sample clock time.)  There are no limitations to the
fringe frequencies that can be tracked with this hardware since accurate
phase for each corresponding data sample is produced.  The station-based
fringe-phase generator equation for the upper sideband case is
\markcite{Thompson1986} (Thompson \etal\ 1986):
\begin{equation}
\label{equation3} \phi(t) =
-2\pi\cdot\tau(t)\cdot\left[ \nu_{LO}-\frac{f_s}{SF}\right],
\end{equation} 
where $\tau(t)$ is the station-based delay model, $\nu_{LO}$ is the
sky frequency that is converted to DC (zero frequency) on tape, $f_s$
is the sample rate, and $SF$ (section~\ref{SFref}) is the sampling
factor.  The negative sign in front of equation~(\ref{equation3})
follows the sign convention chosen for the correlator delay: if the
antenna is between the array reference and the radio source it is a
negative delay.  The lower sideband fringe generator equation is
identical except for a sign reversal in equation~(\ref{equation3}).

\subsubsection{\label{mainFIFO}Alignment and Delay Tracking FIFO} This
is a 256k, 1M, or 4M by 25-bit FIFO for absorbing the short term
frequency variations between the correlator clock and the S2 clock, as
well as for compensating for station delay (section~\ref{DG}). The
choice of size is determined by the cost and need for large delays on
the Station Module. (Note that large delays can also be obtained by
controlling the Playback Terminal, but electronic delays can be
controlled more rapidly.) The 25 bits are used to carry two data and
one data-valid bit for each channel plus the 1 Hz tick. Since all
eight data channels are represented across the 25-bit word-size in the
FIFO, the delay is applied to all eight channels at once. The output
clock of this FIFO is modified by the delay-tracking hardware to track
delay to $\pm0.5$ samples.

\subsubsection{Alignment Determination} At all times during the
correlation, the digital data being played back from the tapes must be
time-aligned exactly ({\em i.e.} within one delay interval). Because the
correlator is station-based, it is convenient to define ``correlator
time'', an arbitrary reference time that is close to the actual time
of the experiment. The 1 Hz ticks in the correlator are
``time-tagged'' with this time. The 1 Hz ticks, derived from the S2
tapes, with the same time-tags are compared and then aligned with the
correlator's 1 Hz ticks. The alignment detection counter determines
the delay between the correlator's 1 Hz tick and the 1 Hz tick from
the S2 after it has been fed through the station FIFO.  The maximum
alignment error is $\pm0.5$ samples of delay. Each S2 is treated the
same way.

\subsubsection{S2 Test Vectors} Test vectors are used for testing
connectivity to the S2. The S2 can be commanded to generate a test
sequence on its output cables. This block checks connectivity by
generating the same sequence on the Station Module board and comparing
it with the received sequence.

\subsubsection{Data Switch and Level Control} The data switch
dynamically connects correlator channels to S2 data streams. Any one of
sixteen data/data-valid bit streams from the S2 can be mapped to any of
the eight correlator channels using this switch. This switch is
necessary because S2 channel mapping depends on the sample rate,
quantization, and channelization chosen at record time. The Level
Control block is used for translating various 1 and 2-bit encoding
schemes to the correlator's internal format.

\subsubsection{Test Vector Generator} This block is used to generate
simulated astronomical signals, complete with Doppler shift and delay
for testing the connectivity and performance of the downstream hardware
(including the delay tracking and fringe stopping hardware). The
performance requirements of the correlator indicate that long sequences
are needed to properly exercise the correlator hardware. However, the
limited size of the test vector RAM makes it necessary to repeat test
vectors on a periodic basis. In order to achieve this, the hardware and
software is designed with a user-definable delay model during one
correlator time-slice and then with the mirror image of the delay model
on the next time-slice. Pseudo-random test sequences (`test vectors')
are generated independently on both time-slices. The sequence then
repeats. The delay model is normally chosen such that there are many
delay steps within one time-slice. The use of the mirror image prevents
zero$^{th}$ order discontinuities in the simulated data that would
otherwise cause delay-tracking loss in the correlated output
(section~\ref{PRC}). Fig.~\ref{fig3} more clearly shows how this is
achieved. The dashed line is the model that was used when the test
vector data was generated (curvature is highly exaggerated), and the
solid line is the model that the correlator uses to track it.


\begin{figure}
\vspace{5cm}

\caption[]{
Test vector delay model.  The model in the
second interval is the mirror image of the model in the first interval
so that no net delay change occurs, allowing test vector sequences to
be repeated indefinitely.  The curvature in the plot is highly
exaggerated.
}
\label{fig3}
\end{figure}


Note that the sign of the fringe-phase rate will change every interval.
In a conventional lag correlator this would cause `smearing' and degradation
of the cross-power spectrum, but because of the way the correlator
implements lag-based fringe stopping, this is not a problem
(see section~\ref{CCA}).
Fig.~\ref{fig4} is an example of the real and imaginary lag components
of a 512 lag test vector run. In this sequence there is high SNR, and
there is both a strong continuum and line component to the data. The
continuum component allows accurate measurement of the residual phase
component of the run and the line component allows all of the lags to be
exercised.


\begin{figure}
\vspace{9cm}

\caption[]{
An example of 512 lag in-phase (top) and
quadrature (bottom) test vector cross-correlation.  The test vectors
are chosen to exercise all lags so that even subtle hardware faults
can easily be found.
}
\label{fig4}
\end{figure}


This test vector capability was originally developed to ensure that
there were no systematic biases resulting from the correlator design.
Initially, a test vector sequence was run through a software simulator
whose systematic biases were closely studied and eliminated. The same
sequence was then run through the correlator hardware and the outputs
compared. This approach, using a simulation, was considered to be a
crucial part of the design process. Operationally, pre-defined test
vector correlator jobs are run once per day, and analyzed to ensure
that all of the correlator hardware is functioning properly.

\subsubsection{\label {ToneExtract} Phase Calibration Tone Extractors}
A train of narrow pulses -- resulting in a `comb' of evenly spaced
sinusoids across the observed band -- are often injected into the
feeds of VLBI antennas for the purpose of checking the stability of
the electrical path from the earliest stages of the signal path to the
recording system or correlator. They are also often used to
inter-calibrate the phases of the baseband channels. In order to
extract data for these purposes, digitized signals containing the
tones are cross-correlated with synthesized tones at the
correlator\footnote{This operation need not be performed at the
correlator -- though this correlator provides the facility to do so.}
in devices called tone extractors. In this design, eight programmable
phase calibration (phase-cal) tone extractors, each with 9-bit phase
generation, are available. Eight extractors were built so there would
be at least one available per channel and 9-bit phase generation was
chosen to minimize false correlation with harmonics induced by coarse
quantization. Each one may be connected to any channel on the input or
output of the FIFO. If connected to the output of the FIFO, the phase
of the injected tone is inherently modified in the FIFO by the station
delay model. The phase of the tone extractor is correspondingly
controlled to compensate for this effect.  Data can be dumped from
these accumulators up to 500 times per second.  Fig.~\ref{fig5} shows
three examples of extracted phase-cals that demonstrate the usefulness
of this facility in the VSOP mission. The software that interprets the
tone-extraction data derives the delay values shown in Fig.~\ref{fig5}
from multiple tones, in this case four.


\begin{figure}
\vspace{17cm}

\caption[]{
Phase-cal tone extractor example plots.  In
(a) the phase-cal data is extracted from the VSOP satellite via two
different tracking stations --- in the first, a hardware bug in the
tracking station causes random and frequent delay jumps, which do not
show up in the phase alone, thus requiring the analysis of delay
derived from multiple tones.  In (b) phase-cal is shown for a perfect
VSOP tracking pass.  In (c) a phase and derived delay plot is shown
for a ground radio telescope with some phase wander in the receiving
system.
}
\label{fig5}
\end{figure}





\subsubsection{Quantizer-Statistics Accumulators} Quantizer-statistics
accumulators permit monitoring the fraction of sampled data in each of
the four digital levels. Eight accumulators are available, each of
which can be connected to any channel on the input or output of the
FIFO. Data can be dumped from these accumulators up to 1000 times per
second. The results are used to derive three quantizer thresholds for
each correlator input.  The thresholds are used in an inversion
algorithm, to convert raw correlation coefficients to normalized
correlation coefficients.  Following the treatment given in Hagen and
Farley (1973), but in addition including all quantizer thresholds from
both stations, the equation relating the actual correlation
coefficient, $\rho$, and the correlator output, $<r4>$, is:
\begin{equation} \label{equation20}
{\begin{array}{l}
<r4(\rho)> = \frac{1}{\pi} \cdot \int\limits_{0}^{\rho} \frac{1}{\sqrt{1 - r^2}}
\cdot \lbrack \frac{n \cdot (n-2)}{2} \cdot (e^{\alpha(v_{0x}^-,v_{0y}^-,r)} +
e^{\alpha(v_{0x}^-,v_{0y}^+,r)} + e^{\alpha(v_{0x}^+,v_{0y}^-,r)} +
e^{\alpha(v_{0x}^+,v_{0y}^+,r)} ) \\
+ n \cdot (e^{\alpha(v_{0x}^-,\Delta_y,r)} + e^{\alpha(\Delta_x,v_{0y}^-,r)} + 
e^{\alpha(\Delta_x,v_{0y}^+,r)} + e^{\alpha(v_{0x}^+,\Delta_y,r)} )
\rbrack dr
\end{array}}
\end{equation}
where $\alpha(v_1, v_2, r) = -\lbrack {{v_1^2 + v_2^2 - 2 \cdot v_1
\cdot v_2} \over {2 \cdot \sigma^2 \cdot (1 - r^2)}} \rbrack$,
$v_{0x}^+$ and $v_{0x}^-$ are the upper and lower quantization levels for
the $X$-station, $v_{0y}^+$ and $v_{0y}^-$ are corresponding levels for
the $Y$-station, $\Delta_x$ and $\Delta_y$ are the respective central
quantization levels, and $r$ is a dummy variable. In this
implementation, $n$ has been assigned the numerical value of 3 in the
product for the regions above and below the extreme quantization
levels, and 1 is the value for the two inner regions. $\sigma^2$ is
the absolute power level of the analog input signal, and since the
power level is unknown at this stage, $\sigma$ is set to 1. After the
correlation coefficients have been normalized in this way, the
dependence on variations in quantizer thresholds is minimized, within
the dynamic range of this method.

\subsubsection{Synthetic Autocorrelator} The `synthetic'
autocorrelator measures the autocorrelation function in a single FPGA
by accumulating lag data in a single accumulator and using a
programmable delay line (dedicated FIFO) to set the appropriate
lag. The correlation is carried out in a time-division multiplexed
fashion. This is an inexpensive way to obtain a good representation of
the autocorrelation spectrum. The autocorrelator can be connected to
any one of the eight correlator channels on the output of the FIFO and
can provide up to 4096 spectral points on each channel.
Autocorrelation for multiple correlator channels is acquired through
multiplexing. An example of a `synthetic' autocorrelation spectrum
obtained this way is shown in Fig.~\ref{fig6}.


\begin{figure}
\vspace{4cm}

\caption[]{
Example of a synthetic autocorrelation
spectrum. The spectrum was obtained by sweeping over 512 lags,
integrating for 10 ms per lag point. This was repeated for 50 times to
provide a total integration time of 256 s. Narrowband interference at
the center of the band is clearly visible.
}
\label{fig6}
\end{figure}

This method of measuring the spectrum assumes that the signal is
statistically stationary during the integration period. Normally the
synthetic autocorrelator is programmed to spend only a short time at
each lag, typically 10 ms. A complete spectrum is obtained as quickly as
possible, and further time-averaging is accomplished by
averaging an ensemble of spectra. This provides some measure of 
protection against distortion if there are non-stationary components
in the signal.

Nevertheless, the autocorrelator is sensitive to bursts of unflagged
errors if those errors create changes in the statistics. Frequently,
this turns out to be tape errors or short bursts of narrow-band
interference. If sequences of non-random errors that autocorrelate at
large lags are present, high peaks will occur unexpectedly --- producing
a very noisy spectrum. At first glance, this appears to be an unwanted
effect, but on the contrary it has unexpectedly proved to be a powerful
diagnostic for systematic problems.

\subsubsection{Pulsar Timers} There are two independently programmable
pulsar timers --- chosen so that two widely different sky frequencies
(each having significantly different pulse arrival times) can be
gated.  Each timer contains a 32-bit pulse phase accumulator and a
32-bit pulse rate register that are initialized with new coefficients
every correlator time-slice and are thereafter free-running to yield a
pulse phase every sample time.  Each timer can be used to form a pulse
gate that uses the validity channel to gate the data of one or more channels
independently or in tandem. More detail is given in
section~\ref{PGFD}.

\subsubsection{Bus Interface} The Station Module contains a 32-bit VSB
bus interface \markcite{VSB1986} (VME Subsystem Bus 1986) for allowing
an external CPU to control all on-board functions and read out all
data.  The VSB bus can accommodate from two to six slots and multiple
VSB busses can be installed in one VME (sub-rack) backplane.  This
architecture offered considerable performance flexibility when it came
to choosing how many Station Modules would be controlled by one
external CPU board --- an important consideration in ensuring that the
real-time response requirements of the Station Module could be met.

\subsubsection{Gbit/s Multiplexers} The data (16 bits), data-valid (8 bits),
phase (8 bits), delay (1 bit), and 1 Hz + 1 kHz frame clock (1 bit)
signals of the Station Module make up 34 parallel bit-streams
clocked at the sample rate. The function of each Gbit/s Multiplexer is
to convert the bit streams to a single serial Gbit/s signal for
connection via Gbit/s switches (Serial Distribution Modules - see
below) to the downstream Baseline Modules. Since data from each
Station Module is now on one or two thin cables, they can easily be
routed through the system.  Converting these data to high-speed serial
format before transmission is a step of vital practical importance
since it greatly reduces wiring complexity, and hence the size and
cost of the entire system. Either one or two Gbit/s Multiplexer units
are installed, depending on the anticipated data-rate requirements of
the system. This circuitry was installed as a daughter module so that
obsolescence of the `single-source' multiplexer chip would not require
a complete re-design of the Station and Baseline Modules.


\subsection{Serial Distribution Module (SDM)} The function of the SDM is
to provide flexible routing of the data from the Station Modules to the
Baseline Modules. The SDM is implemented on a standard size 6U VME card,
and contains a 16-input to 32-output, unidirectional, 1 Gbit/s cross-bar
switch, which is controlled through the VME bus by an external CPU
module. Because of the high data rates on the signal lines of this card,
careful attention to layout and design is necessary. All Gbit/s inputs
and outputs to the switch are carried on $\rm \sim 3~mm$ diameter
`twin-ax' cable. Input connection is made through the front panel using
3 pin, 0.100" twin-ax connectors and each one carries an AC-coupled
differential ECL signal. This method provides a convenient disconnect
point and a compact design, while maintaining control of impedence. The
16x32 cross-bar switch is actually made of two 16x16 cross-bar chips.
Inputs to the module are attenuated through a 3 dB pad, terminated, and
then split into equidistant unterminated transmission lines to the two
chips. This configuration produces balancing reflections such that the
signal at the unterminated inputs of the chips is almost ideal. A 16x32
cross-bar configuration was chosen so that only one SDM module is
required in each baseline subrack, each one being fed by multiple
Station Modules routed through a high speed fanout amplifier.


\subsection{Baseline Module} The Baseline Module (Fig.~\ref{fig7}) is
where final cross-correlation occurs. Each Baseline Module takes the
data, data validity, timing, and models from two Station Modules
(station $X$, and station $Y$), and performs the fine delay (Vernier
Delay) and Doppler shift correction (fringe stopping) before
performing the final cross-correlation. The module is implemented on a
VME standard size 6U x 400 mm circuit board and occupies one VME
slot. It has dual 32-bit bus interfaces, a VMEbus interface and a VSB
bus interface. The board is configured via the VMEbus and correlation
coefficients can be read out on either bus. With a dual bus
configuration it is possible to configure a given system for a range
of performance capabilities based on the dump time and spectral
channel requirements. A low performance system could use one CPU per
rack and readout via the VMEbus whereas a high performance system
could use multiple CPUs in a rack each connected to one or more
Baseline Modules via multiple, independent VSB busses.


\begin{figure}

\vspace{9cm}

\caption[]{
Simplified Baseline Module block diagram.
The main data paths are shown as thick black lines. The P1 and P2
connectors are used for the dual (VME/VSB) bus interfaces and
the system clock (section~\ref{sysclock}).  Data coming from the
high speed serial demultiplexers on the left are synchronized to the
system clock using FIFOs. Synchronized data from the FIFOs are switched
to one of several primary cross-correlators or are routed through the
secondary lag chain. The precomputation delay FIFOs can optionally be
inserted into the data path. The functions of the Vernier Delay, Phase
Modifier, and the Phase Residual Coefficient generator are explained in
the text.}
\label{fig7}
\end{figure}


\subsubsection{Gbit/s Demultiplexers} Data from the $X$ and $Y$
stations enters the Baseline Modules in high speed serial form. The
original data streams are re-created by 1-Gbit/s serial-to-parallel
data demultiplexer modules, one for the $X$ station input and one for
the $Y$ station input.  The output of these modules goes to 512 deep
synchronous FIFO RAM buffers allowing the incoming timing to be phase
synchronized to the on-board timing, which is in turn synchronized to
the System Clock (section~\ref{sysclock}). These FIFOs eliminate the
need for delay-path matching through the Gbit/s distribution and
switching system and permit wide physical separation of modules in a
large system.  The depth of the FIFO RAM buffers and the sample clock rate
determines the amount of delay-path mismatch that can be tolerated
in this configuration.

\subsubsection{$X$ and $Y$ Data Switches} $X$ and $Y$ data switches
allow any of eight $X$-input sample channels to be correlated with any
of eight $Y$-input sample channels.  They also expand the 1-bit
compressed phase streams for each of the 16 input channels to the
4-bit phase streams required for final fringe stopping at each lag.

\subsubsection{\label{VDPM} Vernier Delay and Phase Modifier} The
delay between the sample streams from two stations cannot be matched
perfectly by the insertion of integer-sample delays. Since the
station-based $X$ and $Y$ delays can be corrected only to $\pm0.5$
sample-interval each, the baseline-based difference can be as large as
$\pm1.0$ sample-interval when subtracted on the Baseline Module. (Note
that the subtraction cannot be done on the Station Modules because
both station delays are not available together.) The function of the
Vernier Delay generator is to take the high order bits of the result
of the $X - Y$ delay subtraction and use them to correct the
baseline-based delay to $\pm0.5$ sample-interval. The Vernier Delay
equation is:
\begin{equation} \label{equation1}
              VD = \left\{ \begin{array}{ll}
              +1  & \mbox{if $[(\tau_x-\tau_y) \bmod 1.0] > 0.5$ } \\
              -1  & \mbox{if $[(\tau_x-\tau_y) \bmod 1.0] < -0.5$, } \\
              0  & \mbox{otherwise}
              \end{array}
      \right. 
\end{equation}
where $\tau_x$ and $\tau_y$ are $X$ and $Y$-station delays in samples,
and a +1 delay indicates that one sample-interval of delay is inserted
in the $Y$-station data path.

Because of the residual delay error of $\pm0.5$ samples, there will be
a non-zero phase-slope across the frequency band at any instant
\markcite{Thompson1986} (Thompson \etal\ 1986). Also, because the
phase slope changes during an integration period, there will be an
unavoidable loss of coherence. At a particular baseband frequency the
phase is undergoing a sawtooth function of time -- a sawtooth step
occurring when there is an integral change in baseline delay, the
amplitude of the sawtooth being proportional to baseband frequency. At
zero frequency the amplitude of the sawtooth is zero. The frequency at
which the amplitude of the sawtooth is zero can be shifted by applying
an appropriate sawtooth cancelling function for that frequency,
referred to in this paper as the frequency of zero phase-error. (Only
at this frequency is there no loss of coherence.) The minimum loss of
coherence occurs when the frequency of zero phase-error is at the
center of the band.  This is implemented as a combination of a
station-based phase slope (equation~\ref{equation3}) and phase jumps
produced by the Phase Modifier. The Phase Modifier equation is:
\begin{equation}
\label{equation2}
\phi_{offset} = \frac{2\pi}{SF} round(\tau_x-\tau_y),
\end{equation}
where $SF$ \label{SFref} is defined here as the ``sampling factor''.
$SF = 2 f_s / {\rm B}$, B is the baseband bandwidth, and {\it round()} is
the rounding function to the nearest integral sample of delay. The
frequency of zero phase-error is at $f_s / SF$.  Normally, VLBI data
is Nyquist sampled, and $SF = 4$. However, if oversampled data is
taken, then for two times (2X) oversampling, $ SF = 8$; for 4X
oversampling, $ SF = 16$. Oversampled data is sometimes taken in
spectral line observations where the total bandwidth is small. $SF =
\infty$ produces zero phase-error at zero frequency. If, in
equation~(\ref{equation2}), the Phase Modifier is set so that zero
phase-error occurs at the center of the Nyquist sampled band (i.e.
$SF=4$), $\tau_x$ and $\tau_y$ must contain two bits of integer-delay
information so that the four Phase Modifier states (0, $\pi/2$, $\pi$,
$3\pi/2$) can be stored. The encoding of these states as delay
information is described in section~\ref{DG}.

\subsubsection{\label{PRC} Discrete-step Delay-tracking Correction -- PRC
Generator} As noted above, there will be a loss of coherence due to
rapidly changing phase-slopes across the band at all but the frequency
of zero phase-error. A gain correction for this effect can be calculated
and applied, but the correction also depends upon the data-valid
patterns and delay-model changes that occur within an integration
period. The data-valid streams can contain long, non-random bursts of
errors due to tape-errors modified by `barrel-rolling' in the recorders
\markcite{Wietfeldt1996}(Wietfeldt \etal~1996).

The Phase Residual Coefficient (PRC) generator calculates a 2-point,
complex spectrum of the time series of the discrete-step
delay-tracking function, qualified\footnote{Correlation is performed
only when the data-valid signal is true.}  by the data-valid
products. This data is used to correct for the discrete-step
delay-tracking loss and resulting residual phase slope in a given
integration interval, taking into account arbitrary data valid
patterns or delay models. The equation for the PRC coefficients, $\rm
P(\Delta f)$ is:
\begin{equation}
\label{equation4} {\rm P(\Delta f)} =
\frac{1}{N_{valid}}\cdot\sum_{i=1}^{N_{tot}}DV_i\cdot \
e^{-j2\pi[(\tau_{xi}-\tau_{yi}) \bmod 0.5]\cdot\Delta f},
\end{equation}
where $N_{tot}$ and $N_{valid}$ are the total number of samples and
the number of valid samples, respectively, in an integration time,
$DV_i$ is 0 if the $X$ or $Y$-station sample is not valid and 1 if both
are valid, $\tau_{xi}$ and $\tau_{yi}$ are $X$ and $Y$ station delays
in sample-intervals, and $\Delta f$ is the frequency difference
between the point in the band where ${\rm P}$ is calculated and the
frequency of zero phase-error in units of cycles/sample. Note that
$[(\tau_{xi}-\tau_{yi}) \bmod 0.5]$ is the fractional sample delay
error. The correlator always calculates ${\rm P}$ for $\Delta f$ of
$f_s/4$ and $f_s/8$. The magnitude of ${\rm P}$ indicates the
normalized amplitude loss due to discrete-step delay-tracking at that
frequency point, and the phase of ${\rm P}$ indicates the accumulated
phase offset. A curve is fitted to three points across the frequency
band --- the two frequencies indicated above plus the frequency of
zero phase-error. (The value of ${\rm P}$ at the zero phase-error
frequency is always $1.0 \angle 0$, by definition, and the function is
symmetrical about that point.) For the Nyquist-sampled case, the
frequency where $\Delta f = f_s/4$ is the band edge; the frequency
where $\Delta f = f_s/8$ is half way from the band center to the band
edge. The curve is used to modify the output spectrum of the
correlator, which has been averaged over an integration time.

Normally there are many baseline delay steps in the correlator
integration time, and the phase is uniformly distributed between its
peak-to-peak values at any given frequency. In this case the curve is
a broad $sinc()$ function \markcite{Thompson1986} (Thompson \etal\
1986) and the magnitude of ${\rm P}$ is $\approx0.90$ at $\Delta f =
f_s/4$ and $\approx0.96$ at $\Delta f = f_s/8$.  However, the {\it
precise} value of ${\rm P}$ is modified by the number of valid samples
in the integration time, and the $X$ and $Y$-station delay model --
both of which can be very complicated functions.

The use of the PRC simplifies the system architecture since baseline
delay models (which can be very complex on space baselines) are not
needed for the calculation, and data validity is already taken into
account.  Additionally, the PRC coefficients could be used to perform
fractional sample correction by dumping at a high rate, Fourier
transforming, and then applying the corrections (although this
functionality is not performed in the correlator in real-time) since
PRC coefficients are always saved with each corresponding correlator
data record.

\subsubsection{Precomputation Delay FIFOs} Dual programmable
precomputation delay FIFOs in the $X$ and $Y$ data paths permit a
particular Baseline Module to be part of a much longer equivalent lag
chain by inserting a digital delay offset before correlation. The
offset is adjusted so that each Baseline Module in the chain
correlates at lag values contiguous to the adjacent Baseline
Modules. If these FIFOs are active, the 10 bits of $X$ and $Y$ delay
information are fed through them along with data and phase values,
before being used for Vernier Delay, Phase Modifier, and PRC
generation. Ideally, these corrections would be applied at each lag
value, but it is practical to apply them only at the end of a chain of
lags. By applying them at the end of each lag segment (equivalent to
the lag range on a single Baseline Module), the errors are
limited. The use of precomputation delay has been used in other XF
correlator implementations \markcite{Whitney1982} (Whitney 1982).

\subsubsection{Correlator Lag Modules} Each Baseline Module contains a
Primary and a Secondary Correlator Lag Module (CLM). These are
daughter boards that plug into the main correlator board and contain
only correlator chips. Each CLM contains a array of sixteen correlator
chips, each with sixteen complex lags. If installed in the primary daughter
board position, the CLM can perform up to 16 independent
cross-correlations or can be chained in various combinations to
provide more lag channels for fewer cross-correlations. The Secondary
CLM can be used only with all chips chained together and only if
chained with the primary CLM. Its function is to increase from 256 to
512 the number of complex lag channels that a particular board can
produce\footnote{The number of complex spectral channels is always
half the number of complex lag channels in a cross-correlator.}.
Correlator chips are mounted on daughter modules to ease routing
requirements, and to permit a more compact assembly of circuit boards.

Fig.~\ref{fig8} is an example of the output of the baseline module,
and in fact, the entire correlator.


\begin{figure}
\vspace{19cm}

\caption[]{
A simple example of `sky' fringes produced by
the correlator.  (a) the raw values of real and imaginary amplitude
versus lag for one dump, (b) the Fourier transformed magnitude and
phase versus frequency for the data in (a), and (c) phase versus time
for about 15 minutes of correlator dumps and magnitude versus
frequency, incoherently averaged over the same time period.
}
\label{fig8}
\end{figure}






\subsection{\label{CCA} Correlator Chip Architecture} Each chip is a 16
complex-lag cross-correlator. The chip contains necessary circuitry
for Vernier Delay correction, Phase Modification, Doppler shift
correction (fringe stopping), and cross-correlation.  The correlator
multipliers are 4-level reduced-product-table multipliers\footnote{By
omitting some low level cross-products, multiplier hardware is
significantly reduced, yet there is negligible impact on the result
\markcite{Thompson1986} (Thompson \etal\ 1986).}  that support both 2
and 3 level multiplication.  The accumulators are full 32-bit
non-truncated accumulators.

Unlike `conventional' lag-based VLBI correlator chips, where fringe
rotation is performed on the $X$ or $Y$ station data at one end of the lag
chain, this one carries the phase of the data along with it through
the lag chain and performs fringe rotation at each lag. This technique
permits many correlator chips to be chained without any smearing of
the correlation (see below). This technique adds only slightly to the
size of the chip, while maintaining the basis of simplicity of the XF
correlator design, the `step and repeat' architecture. The method also
inherently removes Doppler shift in the signal due to antenna motion
relative to the chosen reference location (normally the geocenter). The
details are given below.

A data validity counter is provided for each lag, thus permitting
normalization of accumulator data to be carried out for every
lag. Every accumulator is double buffered so that data from the
previous accumulation interval can be read out while correlator
processing continues. Fig.~\ref{fig9} shows the critical functional
architecture of the correlator chip. The chip also contains logic (not
shown) to allow it to be used independently or to be connected to
adjacent chips in a lag chain. 


\begin{figure}
\vspace{16cm}

\caption[]{
Simplified correlator chip architecture
showing the Vernier Delay, Phase Modifier, and the details of fringe
stopping and correlation for each complex-lag.  In (a), the architecture
of an 8-lag chip is shown -- complex-lags are numbered according to the
convention set out in equation (\ref{equation5}).  X-station data and phase
enters a delay line on the right of the figure.  Y-station data and phase
enters another delay line on the left of the figure with the intervening
Vernier Delay and Phase Modifier as shown.  $Z^{-1}$ is the z-transform of the
unit delay (1 sample time). For clarity, logic required to chain multiple
chips is not shown.  In (b), the architecture of a single complex-lag
is illustrated.  The {\it sin} and {\it cos} blocks produce 3-level
approximated outputs which are then mixed with the Y-station data and
correlated with the X-station. Each chip in the implementation contains 16
complex lags.}
\label{fig9}
\end{figure}




By inspection of Fig.~\ref{fig9}, the amplitudes of the in-phase
component of the output of the correlator at a particular lag, $l$,
is:
\begin{equation} \label{equation5}
I_{XY}(\tau_{lx}-\tau_{ly})=\frac{1}{N_{\tau_{lx}-\tau_{ly}}}\cdot \
\sum_{i=1}^{N_{tot}} DV_{t_i-\tau_{lx}}DV_{t_i-\tau_{ly}} \
X_{t_i-\tau_{lx}}Y_{t_i-\tau_{ly}}cos[\phi_x (t_i-\tau_{lx}) \ -\phi_y
(t_i-\tau_{ly})],
\end{equation}
where $\tau_{lx}$ and $\tau_{ly}$ take on positive values of lag
(referred to the $X$ and $Y$ inputs, respectively, of the correlator
delay lines (Fig.~\ref{fig9})), $N_{\tau_{lx}-\tau_{ly}}$ is the
number of valid data samples out of the total $N_{tot}$ at the given
delay, $DV_{t_i-\tau_{lx}}$ is 1 if the data is valid and 0 if it is
not valid, $X_{t_i-\tau_{lx}}$ and $Y_{t_i-\tau_{ly}}$ are the signals
after delay has been removed to within $\pm0.5$ sample intervals
(section~\ref{VDPM}) and take on values of $\pm3$ and $\pm1$ for 4
level correlation and $\pm3$ for 2 level correlation, the $cos$
function ($sin$ for the quadrature version) is approximated with 3
levels \markcite{Thompson1986} (Thompson \etal\ 1986), and $\phi_x
(t_i-\tau_{lx})$ and $\phi_y (t_i-\tau_{ly})$ are the $X$ and $Y$
fringe-rotator phases at the given lag corresponding to $\tau_{lx}$
and $\tau_{ly}$, respectively. These are the {\em model phases} that
were initially computed on the Station Modules (section~\ref{FG}), and
embedded with the data.

Equation~(\ref{equation5}) can be written as an idealized, long-term
average, as shown in Equation~(\ref{equation6}). This equation has
also been simplified by dropping the data valid qualifiers and making
the variable change, $ t_i^\prime = t_i - \tau_{max} / 2 + \tau_l / 2
$, where $\tau_{max}$ is the length of the delay line, and $\tau_l$ is
the relative delay of the samples on the $X$ and $Y$ delay lines,
referred to the center of the delay line. In the more succinct complex
form,
\begin{equation} \label{equation6}
R_{XY}(\tau_l) = I_{XY}(\tau_l) + jQ_{XY}(\tau_l) =
{\rm E}\{z_x(t) \cdot z_y^*(t+\tau_l) 
\cdot e^{-j(\phi_x(t) - \phi_y(t+\tau_l))} \}
\end{equation}
where $R_{XY}$ is the complex output of the correlator, $\rm E\{\}$ is
the expected value over an integration period, $z_x$ and $z_y$ are the
complex random processes representing the input signals from the $X$
and $Y$ stations, respectively, the exponential term is the complex
equivalent of the cosine term in Equation~(\ref{equation5}), and the
prime has been dropped from $t$.

$z_x(t)$ is a bandpass process representing the $X$-station noise
signal centered near $\omega_0$, which can be written in Rice's
representation \markcite{Papoulis1991} (Papoulis 1991) as $z_x(t) =
w_x(t) \cdot e^{j \omega_0 \cdot t} \cdot e^{j \hat{\phi}_x(t)}$,
where $w_x(t)$ is the random modulation process, and $\hat{\phi}_x(t)$
is the phase imposed by the motion of the station relative to the
reference position ({\em i.e.} Doppler shift), and the drift of the
station clock and higher order terms. $\phi_x(t)$ is the model
estimate of $\hat{\phi}_x(t)$. (If the band has been shifted to
baseband by Local Oscillators, then $\omega_0$ is half the sampled
bandwidth, $f_s/4$.) Note that although $w_x(t)$ is assumed to be
stationary over an integration period, $z_x(t)$ is a non-stationary
process, since its autocorrelation, $R_{zz}(t_1, t_2)$, is time
dependent.  Substituting the above relation for $z_x(t)$ and an
analogous one for the $Y$-station into Equation~(\ref{equation6})
yields, after some re-arranging of terms:
\begin{equation}
\label{equation7} R_{XY}(\tau_l) = {\rm E}\{w_x(t) \cdot
w_y^*(t+\tau_l) \cdot e^{-j \omega_0 (t - (t+\tau_l))} \cdot
e^{-j(\phi_x(t) - \hat{\phi}_x(t))} \cdot e^{-j(\phi_y(t+\tau_l) -
\hat{\phi}_y(t+\tau_l))} \}
\end{equation}
If the models accurately predict the station phase functions,
$\hat{\phi}_x(t)$ and $\hat{\phi}_y(t)$, then $R_{XY}(\tau_l) = e^{-j
\omega_0 \cdot \tau_l} \cdot {\rm E}\{w_x(t) \cdot w_y(t+\tau_l) \cdot
e^{-j \delta \phi_{xy}} \}$, where $\delta \phi_{xy}(t) = \phi_x(t) -
\hat{\phi}_x(t) + \phi_y(t) - \hat{\phi}_y(t)$ is a small residual
term. The objective of the correlator is to extract ${\rm E}\{w_x(t)
\cdot w_y(t+\tau_l)\}$. The term $ e^{-j \omega_0 \cdot \tau_l}$ is
actually accounted for as a simple frequency shift in the output
spectrum. Note that this result is independent of the form of
$\phi(t)$, as long its variation is sufficiently slow that sampling
theory is not violated.  Since the phase data is carried along with
the station data, the sampling rate is sufficient to handle Doppler
shifts equal to the bandwidth, at which point correlation is
pointless, since sky frequencies will no longer overlap between
stations. The residual phase term, $\delta \phi_{xy}$, must be
sufficiently well behaved that the concomitant coherence loss does not
fall below the signal-to-noise threshold. Otherwise, fringes will not
be detected at all.

As noted above, phase rotation sometimes is placed at the end of the
delay line.  This practice incurs an error, referred to in this paper
as `smearing', which can be evaluated approximately by expanding
$\phi(t)$ in a Taylor's series about $t$ as a function of $\tau_l$:
$\epsilon_\phi = \omega_f \cdot \tau_l + \frac{1}{2} \dot{\omega}_f
\cdot \tau_l^2 + ...$, where $\omega_f$ is the fringe rate, $\frac{d
\phi}{dt}$. Smearing occurs when changes in phase rate, made at one
end of the lag chain, occur during the time taken for a data-sample to
transit the lag chain, and denotes a lag-dependent loss of
coherence. If $\tau_l$ is very short, as in continuum observations, or
if the fringe rates are slow, as for low frequency ground-based VLBI
observations, then this error is tolerably small. Smearing can also be
corrected by using sufficiently short integration periods, and then
applying a delay-dependent correction to the correlation
coefficients. We have elected to avoid this method.

This mode of fringe stopping (where the above $sin$ and $cos$
functions are approximated by 3 levels and 16 phase steps) was
empirically determined to degrade the SNR by $\approx0.64$\% as a
result of the phase differencing, compared to the conventional XF VLBI
fringe stopper that applies phase shifts only to one of the
signals at one end of the lag chain. The fringe rotation efficiency is:
\begin{equation} \label{equation9}
  \eta (n) = \frac{1+\sqrt{2}}{\sqrt{6}} \cdot
  \frac{sin(\frac{\pi}{8})}{\frac{\pi}{8}} \cdot
  \frac{sin(\frac{\pi}{2^n})}{\frac{\pi}{2^n}},
\end{equation}
where $n$ is the number of phase bits and is set to 4 in the chip
implementation.  The first $\frac{1+\sqrt{2}}{\sqrt{6}} \cdot
\frac{sin(\frac{\pi}{8})}{\frac{\pi}{8}}$ factor is the efficiency for
3 level fringe rotation \markcite{Thompson1986} (Thompson \etal\ 1986),
and the second $sinc$ factor is due to the phase dithering that occurs
when independently rotating quantized phases are differenced.  The
resulting cross-correlation function has a constant phase bias of:
\begin{equation} \label{equation10}
  bias(n) = \pi \cdot \sum_{i=1}^{n-3} 2^{-(3+i)}
\end{equation}
For $n=4$ bits, the efficiency is 0.954302 and the constant phase bias
is $\frac{\pi}{16}$.

Data-valid accumulations are carried out for each lag in the
correlator (Fig.~\ref{fig9}) and used to normalize the summations at
each lag.

The correlator chip is implemented in a Xilinx ``hardwire'' FPGA. The
large accumulators are, for the most part, implemented in on-chip RAM
to allow the equivalent of approximately 30,000 gates of logic to be
packed into the device that is marketed as a 13,000 gate FPGA. The
hardwire process allowed the design to be tested and refined in the
programmable version, and then directly ported, without any design
conversion, test vector requirements, or risk normally associated with
an ASIC (Application Specific Integrated Circuit), to a fixed program
device. This procedure results in a very low ``up-front'' design cost,
known as a non-recurring engineering (NRE) charge. At the same time
the cost per device is also relatively low. The fraction of the chip
area used by the differential phase rotators at each lag is about
11\%. The fraction used by the data-valid accumulators is about 25\%.


\subsection{\label{sysclock} System Clock} The System Clock provides
the correlator with its timebase. It is designed so that a single
signal can synchronize equipment separated over a wide physical
area. The clock is implemented in a small module that plugs into the
back of a VME backplane. The SYS\_CLK signal consists of a 32 MHz
reference clock, an embedded 1 kHz tick, an embedded 1 Hz tick, and a
4-bit sequence count that is incremented for every 1 Hz tick. The TTL
signal is distributed to Station and Baseline Modules on the VME
backplane and is distributed to other subsystems via 50-ohm coaxial
cables. Each additional subsystem contains a backplane driver and
repeater. Each Station and Baseline Module contains a PLL, which
recovers the 32 MHz clock, and circuitry to extract the time ticks and
the sequence count. The time ticks are used for synchronizing events
on the Station and Baseline Modules. The 4-bit sequence count is used
to resolve potential ambiguities between the time-tagging of ticks,
effectively guaranteeing that all subsystems are properly
synchronized.


\section{Software System Architecture} The same basic principles
guiding hardware development were also used in the design of the
system's software.  The software is modular, and can easily be
expanded to support a large distributed correlator system.  All
communication between different functional software modules is
performed using industry standard protocols.  This approach was taken
to minimize real-time bottlenecks\footnote{Also to allow future CPU
performance enhancements with purchased hardware rather than
re-designed hardware.}, and to minimize the amount of hardware and
software, not specifically related to correlator functionality, that
had to be developed.


\subsection{Embedded CPUs} Station Modules and Baseline Modules do
not, themselves, contain any CPUs --- a choice made to eliminate
development of a special real-time operating system with all of the
required features and to ensure that the system is not tightly coupled
to any specific CPU architecture.  Real-time control and configuration
of these modules (and the SDM switch module) is provided by
off-the-shelf, industry standard, CPU modules running the VxWorks
operating system.  Two types of embedded control software are used ---
one for control of Station Modules and one for control of Baseline
Modules.  The CPU that controls Station Modules is the Station Control
CPU (SCC) and the CPU that controls Baseline Modules is the Baseline
Control CPU (BCC). Both CPU types are capable of controlling one or
more SDM switch modules.  All communications to host computers on the
network are via standard TCP/IP or UDP/IP protocols.

The correlator software can be reconfigured to accept changes in the
hardware configuration very quickly.  If more Station or Baseline
Modules need to be added, or more CPUs are added to the network, the
software is configured (by telling a host about the new CPUs and
modules so it can configure the SCCs and BCCs), and the modules are
connected to each other and existing modules via one or more
distributed SDM switch modules.  In this way a new correlator can be
quickly configured, or an existing one can be expanded or modified.

The SCC contains all of the software modules, drivers, and interrupt
handlers for the control of up to five Station Modules. This is a
limitation due primarily to the size of the VSB (a VSB backplane
contains only six slots). It is also a performance limitation since
the CPU can service only five Station Modules every correlator
time-slice ({\em i.e.} interrupt). The SCC also contains the software
necessary to automatically control each S2 that is connected to an
associated Station Module. At system startup, the SCC must be told (by
a host computer) which Station Modules are to be controlled --- the
SCC checks to see if they are available and then builds the required
software modules. Once the software is configured, the SCC accepts
high-level requests from the Host Computer via RPC (Remote Procedure
Call \markcite{Corbin1991} (Corbin 1991)) function calls to start
jobs (see section~\ref{HCCO}), process stations, or establish SDM
connections. Any status
conditions that may need attention are sent as Error, Warning, or Note
messages to the Host Computer for logging, automatic intervention or
possibly operator intervention. Any data captured from the
autocorrelator or data quality accumulators can be saved to local
and/or remote NFS (Network File System) mounted disks or other
devices.

The BCC is similar in concept to the SCC, as are the procedures for
startup. It contains all of the software modules, drivers, and
interrupt handlers for the control of up to 16 Baseline Modules (the
maximum number of Baseline Modules that can be installed in a given
VME backplane).  The BCC can save correlation coefficients to large
local NFS mounted disks, and/or remote NFS devices.

The SCC and BCC software currently runs on CPU boards with a dual
processor architecture: one processor handles all real-time I/O to the
communications ports, allowing the other processor to respond to
correlator interrupts without fail.


\subsection{\label{HCCO} Host Computer and Correlator Operations}
The host computer provides the platform for top-level control of the
correlator.
There must be at least one host computer, but there may be several in a
given system. Each host computer runs one or more tasks (processes) that
function to orchestrate the execution of one or more correlation ``jobs'',
where a job is typically a single VLBI experiment.

There can be up to 8 independent correlator jobs active at any given
time in the system. A correlation job is created by building a set of
batch files (written in a batch language called ``C3L'') and compiling
them into a form ready for execution by the host process. These batch files
contain all information required to process a job, including
definitions of time ranges, channel configurations, baselines, and
cross-correlations that are to be processed. A job is run by an
operator under the control of the graphical user interface (GUI). Once
the job is started, the host process obtains the physical resources (Station
Modules, Baseline Modules) it needs from the Correlator Resource
Server process (a {\it single}, unique process running on any host computer
on the network), builds the necessary high-level requests, and then sends
the requests to the embedded SCCs and BCCs for processing.


\subsection{Graphical User Interface (GUI)} The GUI is the primary
interface to the human operator during correlation.  The GUI displays
messages coming from the embedded CPUs, displays status and time
information, and displays prompts for the operator to install and
remove tapes from S2s as required. The correlator design emphasizes
pre-correlation decision making. Thus, the functionality of the GUI is
very simple and a relatively unskilled operator can run correlation
jobs in response to instructions from the GUI.  The GUI provides the
following simple facilities:

\small
\begin{itemize}
 \item {\bf Start} a job --- signals the Host Computer to begin
       execution of the job from the very beginning.
 \item {\bf Kill} a job --- signals the Host Computer to cease 
       execution of the job and free up all physical resources used 
       by the job.
 \item {\bf Restart} a job at a {\bf specific UT time} --- tells 
       the Host Computer to load a {\em .cxe} file (section~\ref{C3L}) 
       and start job execution at a specific time. This is normally
       used for testing.
 \item {\bf Restart} a job from the {\bf log file} --- tells the 
       host to load in a {\em .cxe} file and a corresponding {\em .log} file and 
       begin execution at an appropriate time. This is used primarily 
       to recover from an unusual exception condition such as a power 
       failure or system crash.
 \item {\bf Kill} specified station or baseline processing ---  
       allows the processing of individual stations or baselines to
       be terminated.
 \item {\bf Change delay and delay rate} --- allows the 
       operator to modify the delay model on a given station in 
       real-time.
 \item {\bf Operator action monitor} window --- contains 
       messages that require specific operator action such as 
       installation or removal of tapes.
 \item {\bf Job monitor window} --- displays the current 
       time, which stations are active and tracking 
       delay, and messages that are 
       generated by the embedded SCC and BCC processors.  In addition 
       to visual monitoring, audio messages are generated when 
       tapes have to be mounted or when there are exceptions 
       that require operator intervention.
\end{itemize}
\normalsize
In addition to the GUI, it is possible to obtain more detailed
information about the correlator and S2s by remotely logging ({\it
rlogin}) into SCCs, BCCs, or S2s and executing some simple
commands.  For example, the automatic control of the S2s can be
manually overridden by fast-forwarding or rewinding the tapes on the
S2 console, but this not normally required.


\subsection{\label{C3L} C3L} C3L is a batch control language that is
specifically designed for programming the operation of the correlator.
The language contains a number of keywords that allow the various
functional blocks of the Station and Baseline Modules to be configured
and operated in a high-level fashion. There are 4 types of C3L batch
files: job files to describe job wide parameters, station files to
describe station parameters, event files to describe time sequenced
station events, and finally, baseline files to explicitly define
baseline correlations (including full as opposed to synthetic
autocorrelations) that are to be performed.  For each correlator job
there is one job and one baseline file, and a station and event file
for each station in the job. The following text is an example of a
complete station file for a simple experiment:

\scriptsize
\begin{verbatim}
station_name = "AT"
event_file = "AT.ev" /* event file for the station */
file_size = 5000
storage_loc = "/usr/local/save"

ch1_out: lo_freqs = 4930.0e6 connect = pt_ch(0,1,0) sideband = USB
         coding = ADC_VLBA channel_name = "CH1_L" valid = TAPE*pgate1

ch2_out: lo_freqs = 4946.0e6 connect = pt_ch(2,3,2) sideband = USB
         coding = ADC_VLBA  channel_name = "CH2_L" valid = TAPE*pgate1

sstat1: connect = ch1_out stat_int = 5.0 valid_count=GATE*PGATE
        save=ON
sstat2: connect = ch2_out stat_int = 5.0 valid_count = GATE*PGATE
        save=ON

tonex_1: connect = ch1_out tone_freqs = 4.0e6
         valid_count = GATE*PGATE tone_int = 5.0 save = ON
tonex_2: connect = ch2_out tone_freqs = 4.0e6
         valid_count = GATE*PGATE tone_int = 5.0 save=ON

auto_corr: auto_lags = 256 auto_int = 1.0 connect=ch1_out,ch2_out
           valid_count = GATE*PGATE,GATE*PGATE save=ON

playback: tape_id = "ATNF0033" sname = "1354-174"
          position = 00:00:00 
          models = "AT_g.sm", "AT_c.sm" /* station model files */
          utstart = 1998-030-15:29:30 utstop = 1998-030-15:52:52
\end{verbatim}
\normalsize

In the example, two channels are active, the sky frequencies are 4930
and 4946 MHz, quantizer statistics accumulation and phase-cal
extraction are turned on, the synthetic autocorrelator is turned on
for 256 lags and a 1.0 second integration time, and there is one tape
playback from 15:29:30 to 15:52:52.\footnote{There are default
values to many variables not shown --- one of these is `quantization'
which defaults to 4 levels.  If 2-level quantization were required,
then the assignment `quantization=2' would have to be in the
ch$n$\_out: field.}

When executed, the C3L compiler reads the batch files,
checks for consistency and errors, and compiles them into a
``correlator executable'' file (this file is XDR-encoded so that it is
transportable across computer
platforms\footnote{XDR is the industry standard eXternal Data
Representation \markcite{Corbin1991} (Corbin 1991)).}).  The compiler
determines when baselines (explicitly defined in the baseline file)
can be processed, given station playbacks starting and stopping at different
times and perhaps on different sources (at different times) in the station
files. Only those playbacks that overlap in time and that are on the same
source will be correlated. This provides considerable flexibility to
automatically and in a single correlator pass perform any defined
correlations (including redundant correlations) between any antennas
pointing at like sources at any times during the experiment.  The correlator
executable file is subsequently used directly by a Host Computer control
process to run the job.

In addition to the C3L compiler, there are several programs that
facilitate reading and collating PCFS (MKIV/PC Field System --
\markcite{Himwich1996} Himwich and Vandenberg 1996) or Space Radio
Telescope (SRT) format log files ({\em slog2cje}), building all of the
skeleton C3L and associated model build files from the collated log
file data ({\em cje2c3l}), and finally, performing an exhaustive
pre-run time consistency check on the correlator executable file 
({\em cxelint})
so that any errors that could occur during correlation are caught
ahead of time. If proper log files are available (or are created from
accurate anecdotal information) it is a very simple procedure to
produce a final executable file that is ready for correlator
execution.


\subsection{Delay Model} The correlator requires an `a priori'
mathematical model of the wavefront delay encountered during an
observation, relative to an agreed reference point. This delay is
inserted as a function of time into the data path for each station
before cross-correlation. Typically, the model computes the motion of
each station-antenna, and any propagation effects that can be
modelled, such as the atmosphere, and in the case of space VLBI,
propagation errors through the space-ground telemetry system. In some
applications it could include motion of the radio source itself. The
delay model may not be complete, but it must be accurate enough to
guarantee that fringes will be within range of the correlator's search
window. Although not strictly a delay model, a model of the clock
offsets for each VLBI station is included in the same way.

In this correlator the model can be as simple as a {\it null} model or
as complex as the model required to describe the motion of a space
antenna and all its known corrections. For ground-based VLBI the delay
model is built from the program, CALC \markcite{Ryan1979} (Ryan and Ma
1979). For space-based antennas, the delay model is built using data
from the NAIF \markcite{Lynch1992} (Lynch 1992) program, which outputs
position and velocity data for the spacecraft, based upon Doppler
navigation measurements. Up to eight model files, each containing a
component of an overall model, can be merged ({\em i.e.} merging
occurs at run-time in the SCC).

The models are prepared in advance of correlation, mainly using
programs that interface to the CALC and NAIF software.  These station
model files may contain either polynomials or delay
points.  Polynomial files are used for most modeling applications and
the software allows up to 7th order polynomials to be defined over a
user-defined time range.  Delay point files are used in the space VLBI
case to correct for the small but relatively rapid variations in delay
resulting from short term delay errors in the link from the tracking
station to the orbiting satellite\footnote{A timing link file is
generated by a tracking station.  This file is used to produce two
station model files: the first one contains polynomial fits (least
squares) to nominally 10 seconds of data; the second contains delay
points that are the {\it residuals} between the polynomial and the
original timing link data --- these delay residuals are typically
about one or two picoseconds in amplitude.}. The delay points are
interpolated to times which are aligned with the 10 ms epochs, so that
a simple linear interpolation can be used for merging with the
polynomial data in the correlator.

At run time, the Station Control CPU reads the polynomial and
delay-point data from the XDR-encoded station model files, and for every
correlator time-slice forms a point-slope interpolation that is fitted
(least square, zero bias) to the merged polynomial and delay-point
data.  These point-slope values are used to set the delay and
fringe-tracking synthesizer hardware to the correct values at that
instant in time. The synthesizer hardware is then free-running until
it is updated for the next correlator time-slice.


\subsection{Correlator Output Data Format} The raw data records
produced by the correlator are saved in XDR-encoded data files. Each
record contains all of the identification information and raw
correlation coefficients required to make it usable for further
processing on a variety of computers. This data is archived to tape
and subsequently `exported' to UVFITS format \markcite{Diamond1997}
(Diamond \etal\ 1997) for external distribution.


\section{Specialized Signal Processing Functions} The correlator
contains several special signal-processing functions that extend its
basic capability beyond a simple VLBI correlator for continuum
observations.  This includes frequency switching for geodetic
applications, pulsar gating, fast (1 ms) dumping, interpolation for
sample rate translation, and `zoom mode' for high resolution
spectral-line processing.


\subsection{Frequency Switching} This function is implemented to support
frequency-switched bandwidth-synthesis observations planned for the
Canadian S2 Geodetic VLBI program \markcite{Petrachenko1993}
(Petrachenko \etal\ 1993). Each correlator channel (up to four will be
used in the Canadian program) can be assigned up to 16 unique states
(where each state corresponds to a different LO frequency) and dwell
times. Each channel can switch through these 16 states in a sequence of
arbitrary length, defined by the user in an external ASCII file. As the
states are sequenced by the correlator, quantizer statistics, phase-cal
tone data, and autocorrelation data is acquired and binned into separate
outputs for each state of each channel. Since raw correlation
coefficients are not binned in this way, the correlator must be dumped
synchronously with each change in state. The correlator data are then
sorted when they are exported to UVFITS.


\subsection{\label {PGFD} Pulsar Gating and Fast Dumping} The correlator
has two independently programmable pulsar timers for each station. Each
one can be programmed with a completely independent epoch (pulse start
time), pulse period, and rate of change of pulse period. The gate ``on
time'' (pulse width) can be programmed to an accuracy of $\pm0.05$\% of
the pulse period. Each timer generates a gate which is used to qualify
(i.e. logically `AND') the appropriate data-valid signals. Use of pulsar
gating to restrict correlation of data only during the time that the pulse
is `on' significantly increases the SNR of the resulting cross-power function.
Pulsar gating can also be used for other applications where it is
desired to blank the data at specific and regular repeating intervals.
One of these applications is in frequency switching, where it is used to
blank the data during the Local Oscillator settling time in the data
acquisition system.

Normally the correlator performs all model updates and coefficient
readouts on 10 ms boundaries synchronized to the 1 Hz UTC tick.
However, it is also possible to perform these functions on 1 ms
boundaries by restricting the available hardware that the embedded
CPUs must access every millisecond.  This is achieved by modifying the
configuration files that the Correlator Resource Server reads when it
boots the system to include only one Baseline Module per embedded CPU.
In this mode, and used in conjunction with pulsar gating (with a
$\leq$25\% duty cycle), it is possible to dump 128 complex-lags per
CPU per ms\footnote{If the integration time is increased, then 128
lags are available for every additional 1 ms of integration time.}.
These lags can be assigned to one complete cross-correlation, or they
can be part of a longer lag chain.  Currently, there are six embedded
baseline CPUs in the correlator -- thus it is possible to dump 768
lags every millisecond.  This fast dump mode is useful for studying
pulse-to-pulse variations of strong pulsars on VLBI baselines
\markcite{Cohen1974} (Cohen and Cronyn 1974) \markcite{Narayan1989}
(Narayan and Goodman 1989) and was first suggested (for use in this
correlator) as a means to de-disperse pulsar data to achieve better
SNR.  Some of the pulsar features that can be studied on VLBI
baselines are integrated pulse profiles, single pulse profiles,
pulse-to-pulse dispersion, pulse-to-pulse spectra, and, by using a
second gate and correlating several passes to effectively achieve $<1$
ms dumping, pulse microstructure.  Fig.~\ref{fig10} is an example of
fast dump pulsar data \markcite{Delrizzo1999} (Del Rizzo 1999).


\begin{figure}[h]
\vspace{5cm}

\caption[]{An example of fast dump pulsar data produced
with 1 ms dumping and a 16 MHz bandwidth.  The pulsar (PSR B1641-45
observed with the Tidbinbilla DSN 70 m and the Australia Telescope
Compact Array) has a pulse period of about 0.45 seconds and a pulse
width of about 20 ms dispersed over 16 MHz of bandwidth.  (a) shows a
contour plot of baseband frequency versus time.  The dispersion of the
pulse with frequency is clearly visible. (b) is a slice at a
particular lag (delay) through the 3-D surface plot of amplitude vs
time and lag.  The single-pulse profile is clear, and, since the gate
was set to 60 ms, the off-pulse noise level is fairly well
established.  It is important to note in (b) that the data has not
been de-dispersed before obtaining the pulse profile, but this is
entirely possible (to within the smear limit imposed by the 1 ms
integration time) since 256 complex frequency channels were acquired
on each dump.}

\label{fig10}
\end{figure}



\begin{figure}[h]
\vspace{9cm}

\caption[]{
The $X$-station signal before and after zero
insertion and the $Y$-station full band signal --- {\bf X1} and {\bf
X2} signals are separately correlated with the like part of the {\bf
Y} signal to obtain a full cross-correlation result (in two parts).}
\label{fig11}
\end{figure}


\begin{figure}
\vspace{12cm}

\caption[]{
4X zoom-mode processing.  (a) is the
original sampled signal.  Since it is sampled at the Nyquist rate,
images of the baseband signal repeat in frequency.  (b) is the signal
after being bandpass filtered by the FIR filter.  (c) is the spectrum
of the decimating function (i.e. the new sample rate, $f_s^\prime$).
Once the filter function is convolved with the decimating function,
the result is shown in (d).  This is a baseband signal at the new
Nyquist rate of $f_s^\prime$.  This signal is then re-quantized
(resulting in an SNR penalty of 12\%) and correlated to yield a result
that has a factor of four improvement in spectral resolution over the
normal unfiltered case.}
\label{fig12}
\end{figure}


\begin{figure}[ht]
\vspace{6cm}

\caption[]{
 32X zoom mode example.  In (a) there are two
maser lines --- the one in the middle of the band is unresolved and
the one at the upper edge of the band is partially resolved.  (b)
shows the line in the middle of the band using a 32X zoom factor with
a displayed bandwidth equal to about the width of the line in (a).
The result in (b) clearly shows detailed structure that otherwise
could not be obtained with available lag hardware.}
\label{fig13}
\end{figure}


\subsection{Interpolation for Sample Rate Translation} This capability
allows data recorded by different stations at different (but related)
sample rates to be cross-correlated. The method performs
zero-insertion interpolation \markcite{Crochiere1983} (Crochiere and
Rabiner 1983) on all sample streams that are less than the maximum
rate so that the resulting sample streams are all at the same sample
rate.  The correlator's fringe rotators are then used to shift the
signal in select sample streams to the correct frequency (in addition
to the fringe rotation required to remove antenna motion Doppler
shift) before cross-correlation.  For example, $X$-station data
recorded in two 8 MHz basebands can be correlated with $Y$-station
data recorded in one 16 MHz baseband.  Both, of course, must originate
from the same band of sky frequencies.  The practical use of this
capability is to permit VLBI observations between observatories which
have developed baseband converter hardware for different bandwidths
and numbers of baseband channels.

Fig.~\ref{fig11} illustrates how this is done.  The $X$-station data
consists of two Nyquist sampled baseband signals that are one-half the
width of the $Y$-station data.  Inserting a zero between every other
sample produces the {\bf X1} and {\bf X2} $X$-station spectra shown in
Fig.~\ref{fig11}.  These must be cross-correlated with the {\bf Y}
signal (shown for clarity as two bands the same width as the X station
bands).  The {\bf X1} signal simply cross-correlates with the {\bf Y}
signal to produce a cross-power spectrum that appears to have
originated from data oversampled by a factor of two.  The {\bf X2}
signal can be cross-correlated with the like part of the band in the
{\bf Y} signal by either shifting the {\bf X2} signal up by half the
band, or by shifting the {\bf Y} signal down by half the band.  This
is done by offsetting the frequency of the $X$ or $Y$-station fringe
rotator by the appropriate amount, independently of the geometric
Doppler model.


\subsection{``Zoom Mode'' for High Resolution Spectral-Line
Processing} The motivation for designing a ``zoom mode'' was to
provide the correlator with very high spectral-line resolution on a
sub-band of the original signal within the constraints of the
available correlator hardware.  Zoom mode signal processing consists
of prefiltering the data through an FIR (Finite Impulse Response)
bandpass filter, decimating it to a lower sample rate, re-quantizing it
(which incurs an additional loss in sensitivity of 12\%) and then
cross-correlating the result to yield a spectrum of the signal that
has a much lower bandwidth, and therefore a much higher spectral
resolution than the same number of correlator channels applied to the
unfiltered signal.  Keeping the design of this mode simple places a
restriction on the characteristics of the bandpass filter --- namely
that the filter passband can be chosen to be only an integer
submultiple of the total bandwidth.  This restriction is necessary
because the process of decimating the filtered signal is what actually
translates the narrow sub-band to baseband.


Fig.~\ref{fig12} illustrates the process for a zoom factor of 4,
choosing the 3rd `frequency-slot' bandpass. Some frequency-slots will
produce a lower sideband spectrum at baseband.  If desired, these can
be converted to upper sideband by using the correlator's fringe
rotator to shift the signal by $\frac{f_s^\prime}{2}$.  The real
advantage of zoom mode becomes evident with zoom factors of 8, 16, and
32, resulting in an improvement in resolution that would be expensive
to achieve by simply increasing the number of correlator lags.  Zoom
mode is especially useful for line observations in the VSOP Space VLBI
mission due to the wide (16 MHz) fixed bands available on the
satellite --- a restriction resulting from the need to minimize the
launch vehicle payload.

The disadvantage of the simple method described above is that it is
possible for the spectral line of interest to fall `between the
cracks' of the available zoom-filter bands and therefore be impossible
to process.  Normally, this is not a problem since the line can be
positioned by offsetting the LOs of the observing antennas so that it
does not fall, for example, at the exact center of the recording band.
Alternatively, it is possible to use a different decimating factor
that allows coverage of these cracks while still staying within the
constraints of the correlator's acceptable clock rates.  For example,
a factor of 25 achieves this but is not currently implemented in the
correlator.

Fig.~\ref{fig13} is an example of how zoom mode has resolved a maser
line with no additional correlator hardware.

\subsubsection{Zoom Mode Implementation} The primary component of zoom
mode is the digital filter that performs the bandpass function.  It is
a 95-tap linear-phase FIR filter implemented in a Xilinx FPGA.  This
FPGA implementation was chosen because it contains a large array of
Configurable Logic Blocks (CLBs) that can be configured as 16xN RAM.
The 16xN RAM is ideal for VLBI data since each RAM block can implement
a 2-bit, 2-tap partial sum of products in a lookup table.  A RAM size
of 16x8 was chosen, which results in 8-bit outputs\footnote{8-bit
quantization was chosen at this stage since the requantization loss is
negligible. In addition, simulation indicated that using more bits did
not yield an improvement in the filter performance.} from the 48 RAM
blocks containing the partial sums of products.  These 48, 8-bit
numbers are summed in an adder tree, the result of which is finally
re-quantized to 2 bits -- incurring a 12\% loss of sensitivity.  The
performance requirements of the adder
tree are reduced significantly since it must operate only at the
decimated sample rate $f_s^\prime$ rather than the original sample
rate $f_s$.  The FIR filter coefficients are generated using the
`Meteor' constraint-based FIR filter design program
\markcite{Steiglitz1992} (Steiglitz \etal\ 1992). The final RAM lookup
table coefficients are prepared, written to files, and made available
for loading by correlator software at run-time as required.


\section{A Summary of the Implemented System \label{builtsystem}} A
modest size correlator system has been constructed at the Dominion
Radio Astrophysical Observatory near Penticton, British Columbia.  It
is fully operational and has the following hardware modules,
configuration, and performance capabilities:

\small
\begin{itemize}

\item There are 10 Station Modules controlled by 2 CPUs in a single
standard VME card cage. Only six Station Modules are currently connected
to S2s (i.e. the system is populated with only six S2-PTs). Each
Station Module contains one zoom FIR filter.

\item There are 48 Baseline Modules controlled by 6 CPUs in 3 standard
 VME card cages.  There are 2 CPUs in each card cage and each CPU has
 VME bus access to all 16 Baseline Modules in its rack and VSB bus
 access to its nearest 4 Baseline Modules.  Each Baseline Module is
 populated with 2 CLMs for a total of 512 complex-lags.  Each Baseline
 Module's primary CLM and data switch is set for a granularity of 32
 complex-lags --- thereby allowing up to 8 cross-correlations of 32
 complex-lags each to be performed.  Baseline Modules can be chained
 in various ways to form longer lag chains but the maximum lag chain
 length is 16384.

\item Each $X$ or $Y$ input of each Baseline Module can be connected
 to any Station Module.

\item Each CPU is connected to its own dedicated 4 Gbyte hard drive
 and the maximum dump rate to each hard drive is about 200
 kbytes/sec\footnote{This is a CPU performance limitation --- the
 Baseline Module design and dual bus architecture allows about a
 factor of 40 better than this.}.  Thus, the maximum aggregate dump
 rate is about 1.2 Mbytes/sec for 5.6 hours.  (The maximum dump rate
 to a remote NFS device is somewhat less and depends on the
 performance of the remote device.)  Each CPU is connected by a
 dedicated 10 Mbps Ethernet\footnote{VxWorks implementation of NFS on
 10 Mbps Ethernet provides a maximum data transfer rate of about 120
 kbytes/sec.}  to a central switch which is connected by a 100 Mbps
 Fast-Ethernet to an Ultra-Sparc computer used for UVFITS export, raw
 data archiving, and data analysis.

\item There is one Sun4 host computer that provides the user interface
 and performs quasi-real-time control of the correlator.

\item The correlator time-slice can be configured for either 1 ms or
 10 ms timing mode.  Normally, 10 ms timing mode is used --- all
 interrupts, delay model, fringe model, and time updates occur every
 10 ms and data can be dumped at most once every 10 ms.  One ms timing
 mode is used only if dumping is to occur at intervals less than 10
 ms.

\item In 10 ms timing mode, all modules are on-line and each CPU can
 dump about 16000 complex-lags/sec if the dump time is $\geq$100 ms,
 and about 33000 complex-lags/sec if the dump time is $<$100
 ms.\footnote{If dump times are $\geq$100 ms, data is XDR encoded and
 normalized as previously described.  If dump times are $<$100 ms, raw
 data is saved and only one data valid counter of the lag chain is
 used to normalize the data, yielding an increase in speed.}

\item In 1 ms timing mode, the system is normally configured so that
 each CPU is dumping data from one Baseline Module.  If $<$25\% duty
 cycle pulsar gating is used, then each CPU can dump 128
 complex-lags/ms.  The system can be configured so that each Baseline
 Module is only a small fraction of a large lag chain.  In this
 configuration up to 768 complex-lags can be dumped every millisecond.
 If no gating is used, each CPU can dump 32 complex-lags/ms and the
 maximum lag-chain length is 192 complex-lags.

\item The correlator can operate at sample rates from 125 kHz to 32
 MHz in doubling increments.  Sample rates below 4 MHz would normally
 be used only in zoom mode after decimation (i.e. since the minimum S2
 data rate is 4 Msample/s). With 8 channels of 32 Msamples/s, 2 bits
 wide, the inherent, overall bandwidth of the correlator, itself, is 512
 Mbit/s (four times wider than the S2 recorder).

\end{itemize}
\normalsize


The authors wish to thank the Canadian Space Agency for providing
funding for the development and operation of the correlator. We also
acknowledge the continued support of the Herzberg Institute of
Astrophysics of the National Research Council of Canada, the Space
Geodynamics Laboratory of the Center for Research in Earth and Space
Technology, and the Geodetic Survey Division of Natural Resources
Canada. Finally, we wish to acknowledge the vision and commitment of
the Institute of Space and Astronautical Science of Japan and agencies
representing all participating ground tracking stations and radio
telescopes for making the VSOP mission, and our contribution to it,
possible.



\end{document}